\documentclass[aps,prl,twocolumn,superscriptaddress,showpacs]{revtex4}
\usepackage{graphicx}% Include figure files
\usepackage{dcolumn}% Align table columns on decimal point
\usepackage{bm}% bold math

\begin{document}
\preprint{APS/123-QED}

\title{Revisiting the valence-band and core-level photoemission spectra of NiO}

\author{M. Taguchi}

\affiliation{Soft X-ray Spectroscopy Lab, RIKEN SPring-8 Center, Sayo, Sayo, Hyogo 679-5148, Japan}

\author{M. Matsunami}

\altaffiliation[Present address:]{ Institute for Solid State Physics, University of Tokyo, Kashiwa, Chiba 277-8581, Japan}

\affiliation{Soft X-ray Spectroscopy Lab, RIKEN SPring-8 Center, Sayo, Sayo, Hyogo 679-5148, Japan}

\author{Y. Ishida}

\affiliation{Soft X-ray Spectroscopy Lab, RIKEN SPring-8 Center, Sayo, Sayo, Hyogo 679-5148, Japan}

\author{R. Eguchi}

\altaffiliation[Present address:]{ Institute for Solid State Physics, University of Tokyo, Kashiwa, Chiba 277-8581, Japan}

\affiliation{Soft X-ray Spectroscopy Lab, RIKEN SPring-8 Center, Sayo, Sayo, Hyogo 679-5148, Japan}

\author{A. Chainani}

\affiliation{Soft X-ray Spectroscopy Lab, RIKEN SPring-8 Center, Sayo, Sayo, Hyogo 679-5148, Japan}
\affiliation{Coherent X-ray Optics Lab, RIKEN SPring-8 Center, Sayo, Sayo, Hyogo 679-5148, Japan}

\author{Y. Takata}

\affiliation{Soft X-ray Spectroscopy Lab, RIKEN SPring-8 Center, Sayo, Sayo, Hyogo 679-5148, Japan}
\affiliation{Coherent X-ray Optics Lab, RIKEN SPring-8 Center, Sayo, Sayo, Hyogo 679-5148, Japan}

\author{M. Yabashi}

\affiliation{Coherent X-ray Optics Lab, RIKEN SPring-8 Center, Sayo, Sayo, Hyogo 679-5148, Japan}
\affiliation{JASRI/SPring-8, Sayo, Sayo, Hyogo 679-5198, Japan}

\author{K. Tamasaku}

\affiliation{Coherent X-ray Optics Lab, RIKEN SPring-8 Center, Sayo, Sayo, Hyogo 679-5148, Japan}

\author{Y. Nishino}

\affiliation{Coherent X-ray Optics Lab, RIKEN SPring-8 Center, Sayo, Sayo, Hyogo 679-5148, Japan}

\author{T. Ishikawa}

\affiliation{Coherent X-ray Optics Lab, RIKEN SPring-8 Center, Sayo, Sayo, Hyogo 679-5148, Japan}
\affiliation{JASRI/SPring-8, Sayo, Sayo, Hyogo 679-5198, Japan}

\author{Y. Senba}

\affiliation{JASRI/SPring-8, Sayo, Sayo, Hyogo 679-5198, Japan}

\author{H. Ohashi}

\affiliation{JASRI/SPring-8, Sayo, Sayo, Hyogo 679-5198, Japan}

\author{S. Shin}

\affiliation{Soft X-ray Spectroscopy Lab, RIKEN SPring-8 Center, Sayo, Sayo, Hyogo 679-5148, Japan}
\affiliation{Institute for Solid State Physics, University of Tokyo, Kashiwa, Chiba 277-8581, Japan}

\date{\today} 

\begin{abstract}
We have re-examined the valence-band (VB) and core-level electronic structure of NiO by means of hard and soft x-ray photoemission spectroscopies. 
The spectral weight of the lowest energy state was found to be enhanced in the bulk sensitive Ni $2p$ core-level spectrum. 
A configuration-interaction model including a bound state screening has shown agreements with the core-level spectrum, and  off- and on-resonance VB spectra. These results identify the lowest energy states in the core-level and VB spectra as the Zhang-Rice (ZR) doublet bound states, consistent with the spin-fermion model and recent $ab$ $initio$ calculations within dynamical mean-field theory (LDA+DMFT). The results indicate that the ZR character first ionization (the lowest hole-addition) states are responsible for transport properties in NiO and doped NiO.

\end{abstract}


\maketitle
NiO has played a very important role in clarifying the electronic structure of transition metal (TM) oxides. 
It has set a framework for understanding the rich physical properties of  TM compounds. 
More recently, the discoveries of (i) bistable resistance switching\cite{seo04} (ii) a giant low frequency dielectric constant\cite{wu02} and (iii) the direct correlation between exchange bias and interfacial spins\cite{ohl03} have led to a major resurgence of interest in NiO, as well as doped and deficient NiO. In order to elucidate the transport properties of NiO and doped or deficient NiO, one must understand the character of holes that enter the acceptor levels or the top of the valence band (VB). 
For a long time, NiO was thought to be a prototype Mott insulator in which the insulating gap is caused by 
on-site Coulomb energy $U(\equiv U_{dd})$. Zaanen, Sawatzky and Allen redefined NiO as a charge-transfer (CT) insulator\cite{zaa85}. New theoretical concepts coupled with photoemission spectroscopy (PES)\cite{oh82,fuj84,saw84,she90,huf94} played a pivotal role in setting this stage. Motivated by resonant VB PES\cite{oh82}, Fujimori and Minami originally applied a configuration-interaction (CI) model to NiO\cite{fuj84}. It was shown that the first ionization state is of mainly $3d^8\underline{L}$ character and $3d^7$-like states occur at a higher binding energy\cite{fuj84}.   
This picture indicated that the O $2p$ band appears between the lower and upper Hubbard bands, and the insulating gap is formed between the 
O $2p$ and the upper Hubbard $d$ bands. 
This approach is well established for explaining core-level PES of TM compounds in general. 
However, for NiO itself, the double peak structure of the lowest energy states in the Ni $2p$ core-level PES 
cannot be explained by a single-site NiO$_6$ CI model. 
This shortcoming was overcome by using a multisite cluster (MSC) model\cite{Vee93prl},
which includes nonlocal screening effects from neighboring Ni sites. It was shown that an extra peak appears at the high binding energy side of the main line due to the nonlocal screening. 
Because the structures seen in core-level spectra arise from energetically different screening channels in final states, this extra intersite screening channel should also be important for other spectroscopies, and particularly so for the VB PES. 
In fact, the similarity in the spectral shapes of the VB and the core-level is well-known\cite{huf94}. Nonetheless, the importance of the nonlocal screening in VB PES of NiO based on a MSC model has not been explored to date. 
Theoretical studies of the VB electronic structure have shown that the two lowest energy  electron removal states are Zhang-Rice (ZR) doublet bound states\cite{elp92,bal,ren06,kun07,yin07}. They arise from a competition between O $2p$-Ni $3d$ hybridization and the Ni on-site Coulomb interaction. These results are consistent with angle resolved PES of NiO, which has shown nondispersive features for two lowest energy states\cite{she90}.
 While, for high T$_c$ cuprates, the relation between the VB and core-level PES has been studied extensively in the context of the ZR singlet bound state, it has received less attention in NiO. 
This is partly because VB PES directly probes the electron removal spectral function ($d^{n}$$ \rightarrow$$ d^{n-1}$) while the core-level PES indirectly probes  many body effects in the presence of a core-hole ($2p^63d^{n}$$\rightarrow$$2p^53d^{n}$). This makes calculations much more difficult for NiO ($d^{8}$$ \rightarrow$$ d^{7}$), in the presence of  multiplets and many body effects, as compared to the cuprates ($d^{9}$$ \rightarrow$$ d^{8}$). Therefore, an explanation of the spectral shapes and its relation to the ZR doublet bound states in both the core-level and the VB, using a unified model, remains an open question. The main purpose of the present letter is to propose an explanation for both core-level and VB PES  based on the CI model, in which the charge transfer from the bound state is included within a single-impurity framework.  The model retains the simplicity and generality of the single cluster approach, while taking into account the screening effects due to the bound states in an approximate way.

Hard x-ray (HX) PES  is a bulk sensitive probe of the electronic structure due to its ability to overcome high-surface sensitivity of conventional PES.  Unlike soft x-ray (SX) PES, TM $2p$ core-level HX PES have shown additional well-screened features with significant intensity at the low binding energy side of the main peak\cite{hor04}. It should be noted that these spectral changes are seen only in the metallic phase, and the features are weak or missing in the insulating phase. 
These features were explained well by the CI model including a screening channel  derived from coherent states at Fermi energy. 
Recent theoretical studies of the MSC model\cite{oka05,vee06} offer an alternative interpretation on the basis of a nonlocal screening scenario. The question whether such features reflect the screening from the coherent states, or the nonlocal screening of the MSC model, and whether
these models can be effectively mapped into each other or not, remains to be answered in general. However, at least in high $T_c$ cuprates, the MSC model has confirmed that the screening from the ZR singlet bound state at the top of the VB is the origin of the lower binding energy feature in the Cu $2p$ core-level HX PES\cite{oka05}. Because NiO is a good insulator, a big change in  HX PES is not expected. However, the nonlocal screening effect was originally proposed for Ni $2p$ core-level PES of NiO. We, therefore, felt it worthwhile to study it by HX PES. This is another motivation of the present letter.

\begin{figure}
\includegraphics[scale=.41]{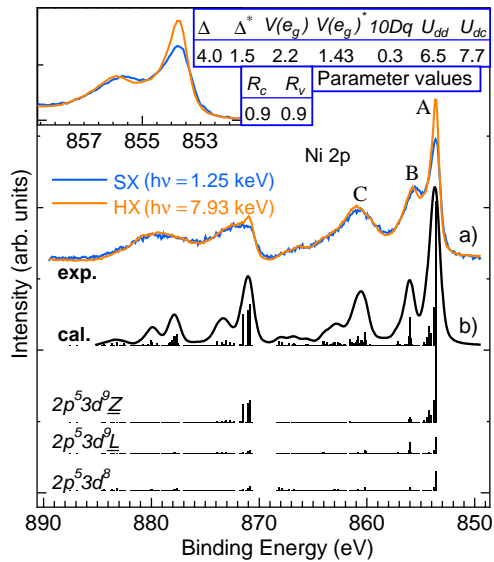}
\caption{\label{fig1}   %FIG.1
(Color online)  (a) Comparison between experimental Ni $2p$ core-level HX and SX PES. (b) A CI calculation for Ni $2p$ spectrum. Final-state components are also displayed. The core-hole lifetime broadening is taken as 0.2 eV. The Gaussian broadening is $0.2 $ eV (half width at half maximum).}
\end{figure}

HX PES measurements were performed 
in a vacuum of 1 $\times$ 10$^{-10}$ Torr
at undulator beam line BL29XU, SPring-8 using a Scienta R4000-10kV electron analyzer. The instrumentation details are described in Ref.~\cite{Tam01}. The energy width of incident x-ray was 70 meV, and the total energy resolution, $\Delta$E was set to $\sim$ 0.2 eV. 
SX-PES and X-ray absorption spectroscopy (XAS) (total electron yield mode) were performed at BL17SU, with $\Delta E \sim$  0.2 eV. All PES measurements used normal emission geometry to maximize depth sensitivity.
Sample temperature was controlled to $\pm 2 $K during measurements. A single crystal of NiO was fractured $in$-$situ$. HX (SX) PES were measured at 200 K (room temperature) to avoid charging.  The Fermi level of gold was measured to calibrate the energy scale. The O $1s$ spectra were clean single peaks, as is known for cleaved NiO.
Core-level spectra were corrected by subtracting a Shirley-type background.

First, we present experimental Ni $2p$ core-level HX PES and SX-PES of NiO obtained with photon energies of $h\nu$=7.93 keV and 1.25 keV in Fig.~1(a). The kinetic energy of the Ni $2p$ core-level corresponds to a probing depth of $\sim$60 $\text{\AA}$ for HX PES as determined by the inelastic mean free path, while it corresponds to $\sim$15 $\text{\AA}$ for SX-PES\cite{NIST}. 
The core-level spectra were normalized at the feature C, since it is well known that the feature C is insensitive to the surface\cite{par99,ald96}.
 In comparison with SX-PES, the intensity of the feature A for bulk sensitive HX PES was found to be enhanced by a factor of 1.35 with respect to the SX-PES (see inset). 
In contrast, negligible change was seen in the features B and C. 
In the MSC model, the most important result is that the feature B originates from the nonlocal screening. 
It was also predicted that nonlocal screening in the peak B is more important for the surface than for the bulk. 
However, our experiments in Fig.~1(a) clearly show a different behavior: only the feature A was appreciably changed
between the HX and SX PES. It should be noted that the present SX-PES result shows the same qualitative behavior as previous SX-PES results\cite{par99,sor07}, in which the intensity of feature A is significantly reduced at nearly grazing take-off angle $i.e.$ in surface sensitive geometry. This clear enhancement of the feature A is therefore a direct consequence of the bulk sensitivity of HX PES. 

\begin{figure}
\includegraphics[scale=.35]{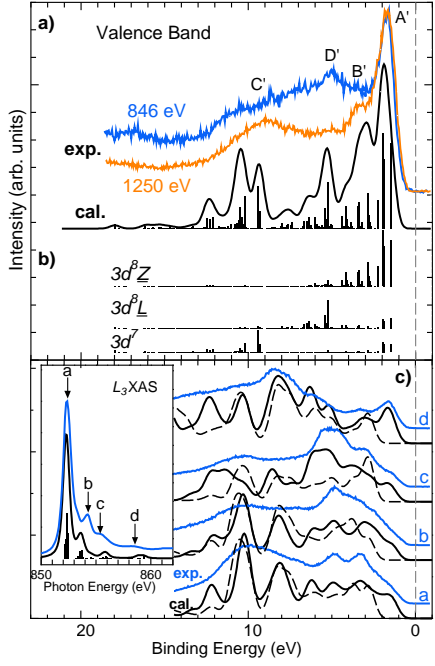}
\caption{\label{fig2}     %FIG.2
(Color online) (a) Experimental and calculated VB spectra with photon energies $h\nu$$=$$1.25$ keV and 846 eV. (b) Final state components for the VB spectrum. (c) Experimental and calculated resonant PES with (thick solid line) and without (dashed line) the bound state screening. Excitation energies are labeled on the $L_3$ edge XAS (inset). }
\end{figure}

In order to address the origin of the observed behavior, we used a CI model with full multiplets in the $O_h$ local symmetry and include a bound state to describe the spectra. The basic framework of our model is as follows: (i) we have used a single-site NiO$_6$ cluster with the usual screening channel for the charge transfer between the Ni site and the O sites. 
(ii) based on results of the spin-fermion model\cite{bal} and the LDA+DMFT calculation\cite{kun07}, we have introduced charge transfer between the Ni site and the bound state within the CI model. The ground state is described by a linear combination of $3d^8$, $3d^9\underline{L}$, $3d^{10} \underline{L}^2$, $3d^9 \underline{Z}$, $3d^{10} \underline{L}\underline{Z}$, and $3d^{10} \underline{Z}^2$,  where $\underline{Z}$ and $\underline{L}$ represent a hole in the bound state and a hole in O $2p$ ligand states, respectively. The Hamiltonian is given by $H$$=$$H_{I}$$+$$H_{II}$. The $H_I$ represents the standard CI model, 
$ H_{I}$$=$$\sum_{\Gamma\sigma} \varepsilon_{3d}(\Gamma)d^{\dagger}_{\Gamma\sigma}d_{\Gamma\sigma} + \sum_{m\sigma}\varepsilon_{2p}p^{\dagger}_{m \sigma}p_{m \sigma} + \sum_{\Gamma\sigma}\varepsilon_{p}(\Gamma)a^{\dagger}_{\Gamma\sigma}a_{\Gamma\sigma}$$+$$\sum_{\Gamma\sigma}V(\Gamma)(d^{\dagger}_{\Gamma\sigma}a_{\Gamma\sigma}$$+$$a^{\dagger}_{\Gamma\sigma}d_{\Gamma\sigma})$$+$$H_{\rm{mult}}$
$+U\sum_{(\Gamma\sigma)\neq(\Gamma'\sigma')}d^{\dagger}_{\Gamma\sigma}d_{\Gamma\sigma}d^{\dagger}_{\Gamma'\sigma'}d_{\Gamma'\sigma'}$$-U_{dc}\sum_{\Gamma m\sigma\sigma'}d^{\dagger}_{\Gamma\sigma}d_{\Gamma\sigma}$
$ \times 
(1 - p^{\dagger}_{m\sigma'}p_{m\sigma'}) $. Here $\varepsilon_{3d}(\Gamma)$, $\varepsilon_{2p}$ and $\varepsilon_{p}(\Gamma)$ represent the energies of Ni $3d$, Ni $2p$ and O $2p$ states, respectively, with the irreducible representation ($\Gamma$=$e_g$ and $t_{2g}$) of the $O_{h}$ symmetry. The indices $m$ and $\sigma$ are the orbital and the spin states. $V(\Gamma)$, $U$, and $-U_{dc}$ are the hybridization between Ni $3d$ and O $2p$ states, the on-site repulsive Coulomb interaction between Ni $3d$ states and the attractive $2p$ core-hole potential, respectively. The Hamiltonian $H_{\rm{mult}}$ describes the intra-atomic multiplet coupling between Ni $3d$ states and that between Ni $3d$ and Ni $2p$ states. The spin-orbit interactions for Ni $2p$ and $3d$ states are also included in $H_{\rm{mult}}$. In addition to the usual CI model ($H_I$ term), we have introduced the bound state labeled '$Z$' as a new screening channel described by $H_{II}$ term, $H_{II} =  \sum_{\Gamma\sigma} \varepsilon_{c}(\Gamma)c^{\dagger}_{\Gamma\sigma}c_{\Gamma\sigma} + \sum_{\Gamma\sigma}V^*(\Gamma)(d^{\dagger}_{\Gamma\sigma}c_{\Gamma\sigma} + c^{\dagger}_{\Gamma\sigma}d_{\Gamma\sigma})$.
An effective coupling parameter $V^* (\Gamma)$ for describing the interaction strength between Ni $3d$ orbitals and the bound state is introduced, analogous to the hybridization $V(\Gamma)$. The CT energy from the bound state to the Ni $3d$ orbital is $\Delta^*$, whereas the usual CT energy $\Delta$ (from the O $2p$ states to the Ni $3d$ states) is defined as the energy difference of the configuration-averaged energies $E(3d^9\underline{L})$$-$$E(3d^8)$. 
The parameter values (in eV) are summarized in Fig.~1.  The Slater integrals and the spin orbit coupling constants are calculated by Hartree-Fock code\cite{cow} and the Slater integrals are scaled down to 85\%.

The calculated spectrum is shown in Fig.~1(b). 
The experimental features are reproduced well by the calculation over the whole energy range. 
To clarify the peak assignment, we show in the lower panel of Fig.~1 the various final state components of the core-level spectra. It is clear that the dominant spectral weight in the features A arises from the $2p^53d^9 \underline{Z}$ state, while the contribution of the $2p^53d^9\underline{L}$ is dominant for the feature B. The satellite C is mainly due to the $2p^53d^8$ state.  The present assignment is different from that by the MSC model\cite{Vee93prl} in which the peak A arises from the $2p^53d^9\underline{L}$ and the peak B is due to the nonlocal screening. 

In order to obtain a better understanding of the bound state screening, 
we applied the present model to VB PES. Figure~2(a) shows the calculated VB spectrum compared to the measured spectra with $h\nu$=1.25 keV and 846 eV. Our 1.25 keV spectrum is consistent with previous reports\cite{oh82,huf94,par99}. The VB PES has a spectral shape quite similar to that of the Ni $2p_{3/2}$ core-level spectrum. 
The only difference between the VB and the core-level spectra is the feature D$'$ (which was enhanced in the 846 eV spectrum).
 The character of the features A$'$-D$'$ can be seen in more detail in Fig.~2(b). One can see that the features A$'$ and B$'$
are dominated by the $3d^8\underline{Z}$ state, with weaker contributions from the $3d^8\underline{L}$ state.
The feature C$'$ is dominated by the $3d^7$ state. The feature D$'$ has mainly $3d^8\underline{L}$ character. 
The most important point
is that two lowest energy electron removal states mainly arise from the "bound state screening" $3d^8\underline{Z}$ state. We recall here that the standard single site CI model for the VB spectra assigns both the features A$'$ and B$'$ to the $3d^8\underline{L}$ state. 
Further, because the MSC is not yet reported for the VB PES of NiO, the present model alone reproduces both the core-level and VB spectra for a common set of parameters. Our results indicate that the feature A$'$ and B$'$ originate from the ZR doublet bound state. 
This indicates that the transport properties of the resistance switching observed in NiO and deficient NiO involves the ZR bound states, which has a composite character. This is because the first ionization state can be reached by simply removing one electron in VB PES. Two theoretical studies have identified the ZR feature for the first ionization state\cite{kun07,yin07}. In fact, one study has generalized it to the ZR bands for the first ionization state in CT insulating TM monoxides from MnO to CuO, as well as layered high-Tc cuprates\cite{yin07}. 

The bound state  screening mechanism is further confirmed by resonant PES at the Ni $L_3$ edge excitation. In Fig.~2 (c) we show the theoretical and experimental resonant PES for incident photon energies labeled a-d in the XAS shown in the inset. The off-resonant spectrum ($h\nu$=846 eV) was discussed above in
comparison with the
$h\nu$=1.25 keV spectrum. Our on-resonant spectra are very consistent with a previous study\cite{tje96}.  The resonant spectra were calculated with (thick solid line) and without (dashed line)
the bound state screening. The agreement between the theory including the bound state screening and the experiments is excellent  for the energy dependence of the spectra. In particular, the spectral intensity changes of the two lowest energy features A$'$ and B$'$, corresponding to the ZR doublet $x^{2}$$-$$y^{2}$ and $3z^{2}$$-$$r^{2}$ bound states, match nicely with the calculations. 

\begin{figure}
\includegraphics[scale=.39]{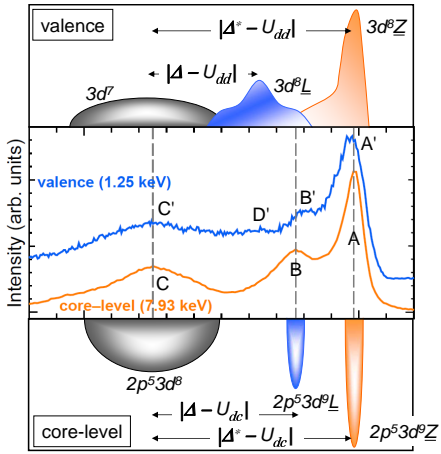}
\caption{\label{fig3}     %FIG.3
(Color online) Schematic of the effect of final states for core-level $2p$ PES and VB PES}
\end{figure}

Finally, we discuss the similarity in the VB and the core-level PES. 
Figure~3 (middle panel) shows the Ni $2p_{3/2}$ core-level HX PES and the 1.25 keV VB PES on the common energy scale. As is well known, the spectral shape presents common features\cite{huf94}. 
In the final state of the core-level PES, a core-hole is created, and the energy of $2p^53d^9\underline{L}$ and $2p^53d^9\underline{Z}$ states are pulled down by the attractive core-hole potential $-U_{dc}$ with respect to that of $2p^53d^8$ states. Because $\Delta^*$$<$$\Delta$, the $2p^53d^9\underline{Z}$ state lies below the $2p^53d^9\underline{L}$ state. In the VB PES, a $3d$ electron is removed in the final state. As shown in the upper panel of Fig.~3, the energy difference between final states are  $E(3d^8\underline{L})$$-$$E(3d^7)$=$\Delta$$-$$U$ and $E(3d^8\underline{Z})$$-$$E(3d^7)$=$\Delta^*$$-$$U$, respectively.  As a result, the $3d^8\underline{L}$ and the $3d^8\underline{Z}$ states are lower in energy than the $3d^7$ state. Again, the $3d^8\underline{Z}$ state lies below the $3d^8\underline{L}$ because $\Delta^*$$<$$\Delta$. 
Therefore, the assignment of the feature A and A$'$, as well as C and C$'$, are the same. But the assignment for the B and B$'$ are not the same. The feature B originates in the $2p^53d^9\underline{L}$ while the feature B$'$ originates in the $3d^8\underline{Z}$ and the feature D$'$ is the equivalent of the feature B (see Fig.~2(a)). This difference is understood in terms of larger multiplet splitting in VB PES leading to the bound state features A$'$ and B$'$, while the smaller multiplet splitting in core-level result in a narrow peak A.

In conclusion, by using the extended CI model, we found an improved agreement for the Ni $2p$ core-level, VB and resonant PES of NiO with a common parameter set. All features in the VB and the core-level PES were reproduced well by including the bound state screening effects, which is beyond the standard CI model. For NiO, the charge transfer effects from the ZR doublet bound states determine the first ionization state (i.e. the lowest hole-addition state). 
We believe that our results provide key features to understand the conduction mechanism of NiO and doped NiO. 
The present study provides the most complete picture to describe the core-level and the VB electronic structure of NiO, consistent with the spin-fermion model and the LDA+DMFT results.

M. T. would like to thanks G. A. Sawatzky and A. Fujimori for useful discussions and illuminating comments.

\end{document}